\documentclass[aps,prd,eqsecnum,11pt,showpacs,preprintnumbers,nofootinbib,superscriptaddress,longbibliography]{revtex4-2}
\usepackage{geometry}                
\usepackage[pdftex]{graphicx}
\usepackage{float}
\usepackage{amssymb,amsmath,amsopn,bm,dsfont}
\DeclareMathAlphabet{\matheul}{U}{eus}{m}{n}
\usepackage{setspace,enumitem}

\allowdisplaybreaks

\usepackage[colorlinks,pdfstartview=FitH]{hyperref}
\hypersetup{linkcolor=blue,citecolor=blue,filecolor=black,urlcolor=blue}

\begin{document}

\title{Simple Bianchi cosmologies with Anisotropic Segre [1(11,1)] dark energy} 
\author{Philip Beltracchi}
\email{philipbeltracchi@gmail.com}
\affiliation{Independent Researcher\\Parker Colorado 80138, USA}

\begin{abstract}
\noindent  In this paper, we use a sample anisotropic but spatially homogeneous metric to look at the properties of Segre type [1(11,1)] universes, that behave like vacuum energy along a spatial plane (two stresses equal to the negative density) but have distinct stress along the third spatial axis. We examine situations with simple relationships between the distinct and degenerate stresses, finding a closed form parameterizations for the case where the distinct stress is equal to zero and finding new conditions in which the case with all stresses equal has a closed form parameterization. We also use numerical methods to illustrate possible behaviors for the case where the distinct stress is the negative of the degenerate stresses.
\end{abstract}

\begin{description}
\item[PACS numbers]
\end{description}
\maketitle

\section{Introduction}
The most common dark energy models are the vacuum energy/ cosmological constant Segre type [(111,1)] models, for which all eigenvalues of the energy momentum tensor are equal, and the perfect fluid [(111),1] models, for which the three eigenvalues with spacelike eigenvectors are equal and the eigenvalue with a timelike eigenvector is distinct. These models are the most popular because they satisfy the spatial isotropy condition commonly assumed in cosmology. See \cite{2012ApSS.342..155B,2021Univ....7..163M,2013CQGra..30u4003T, quintessence2} for for review articles discussing types of isotropic perfect fluid dark energy in cosmology.
However, there are some models of dark energy which involve anisotropic stress. The most general such possible models have Segre type [111,1], or four distinct eigenvalues, see e.g. \cite{2010GReGr..42..763A}. Segre type [(11)(1,1)] models, which behave like vacuum energy along a preferred axis, have also been studied \cite{Dymnikova:2000zi,2023arXiv230109204B}. Another possibility which behaves like vacuum energy in a spatial plane\footnote{More specifically, the principle stresses along a particular spatial plane follow $p=-\rho$, and the energy-momentum tensor and Ricci Tensors are invariant with respect to boosts within that spatial plane and rotations about the third spatial axis, these symmetries are called a ``three dimensional Lorentz group" in \cite{Stephani:2003tm}. The behavior for Segre[(11)(1,1)] is that $p=-\rho$ applies along and axis and the energy-momentum and Ricci tensors are invariant with respect to boosts along and rotations about that same axis.}, which does not seem to have gotten much attention, is [1(11,1)]. In this note, we derive the conditions in which a particular model cosmology degenerates into Segre type [1(11,1)], and further examine some cases in which the distinct spatial eigenvalue is specified by an equation of state.
\section{Derivation and basic geometry}
\label{derivation section}
As was the case with the Segre [(11)(1,1)] systems studied in \cite{2023arXiv230109204B}, the spatial part of a Segre [1(11,1)] tensor has two axes associated with degenerate eigenvalues and the third with distinct eigenvalue, the difference being whether it is the distinct or degenerate ``spacelike" (more properly eigenvalue corresponding to a spacelike eigenvector) eigenvalue which is degenerate with the ``timelike" eigenvalue. To obtain homogeneity/ translation invariance for a cosmology, as well as the invariance of rotation about a preferred axis, we use the same FRLWesque precursor metric as \cite{2023arXiv230109204B}
with the line element 
\begin{align}
    ds^2=-dt^2+a(t)^2dz^2+b(t)^2\Big[\frac{dr^2}{1-k r^2/2 } +r^2 d\theta^2\Big]. \label{met}
\end{align}
This four dimensional spacetime is composed of a two dimensional space of constant curvature\footnote{Eq.(\ref{met}) is written such that $k$ is the Ricci scalar of the two dimensional metric formed by the term in square brackets.} $k$ and scale factor $b(t)$ in polar $r,\theta$ coordinates, a perpendicular spacelike axis with scale factor $a(t)$ and coordinate $z$, and a time axis. This metric is Bianchi type I when $k=0$ and Bianchi type III otherwise, see appendix B of \cite{2023arXiv230109204B} for the derivation and \cite{1898MMFSI..11..267B,1968JMP.....9..497E,Ellis:1968vb,Wald:1984rg,Krasinski:2003zzb,Ellis:2006ba} for information about the Bianchi classification scheme. 

We restate some important quantities for the metric (\ref{met}) which were derived in \cite{2023arXiv230109204B}. One can compute tensor components in an orthonormal frame with the tetrad
 \begin{align}
     e^\alpha_{~\hat{\alpha}}=\left(
\begin{array}{cccc}
 1 & 0 & 0 & 0 \\
 0 & 1/a & 0 & 0 \\
 0 & 0 & \frac{\sqrt{1-k r^2/2 }}{b} & 0 \\
0 & 0 & 0 &
   \frac{1}{r b } \\
\end{array}
\right) .
\label{tet}
 \end{align}
Here the spacetime index $\alpha$ labels the rows and the orthonormal index $\hat{\alpha}$ labels the columns.  The metric in this orthornormal frame is
\begin{align}
g_{\hat{\alpha}\hat{\beta}} = g_{\alpha\beta} e^\alpha_{~\hat{\alpha}} e^\beta_{~\hat{\beta}} = diag( - 1,  1,  1,  1).
\end{align}

The nonzero principle components of the Riemann tensor in the orthonormal frame are
\begin{subequations}
\begin{align}
    R_{\hat{0}\hat{1}\hat{0}\hat{1}}=\frac{-\ddot{a}}{a},\\
    R_{\hat{0}\hat{2}\hat{0}\hat{2}}=R_{\hat{0}\hat{3}\hat{0}\hat{3}}=\frac{-\ddot{b}}{b},\\
    R_{\hat{2}\hat{3}\hat{2}\hat{3}}=\frac{k+2\dot{b}^2}{2b^2},\\
     R_{\hat{3}\hat{1}\hat{3}\hat{1}}= R_{\hat{1}\hat{2}\hat{1}\hat{2}}=\frac{\dot{a}\dot{b}}{ab}.
\end{align}
\label{ReimannA}
\end{subequations}
Dots indicate time derivatives throughout this paper. This has six total components out of a possible twenty and there are two degeneracies, so there are effectively four functions specifying the orthonormal Riemann components. The Weyl behavior can be described with the Q matrix \cite{Stephani:2003tm}
\begin{subequations}
\begin{align}
    Q_{11}=E_{11}=W_{\hat{2}\hat{3}\hat{2}\hat{3}}=\frac{a \left(k-2 b \ddot{b}+2 \dot{b}^2\right)+2 b \left(b \ddot{a}-\dot{a} \dot{b}\right)}{6 a b^2},\\
    Q_{22}=Q_{33}=-Q_{11}/2,
\end{align}
\label{QA}
\end{subequations}
all other $Q$ matrix components vanish, such that it is effectively dependent on a single function.
The Ricci and Kretchman scalars are
\begin{align}
    \mathcal{R}=\frac{a \left(4 b \ddot{b}+2 \dot{b}^2+k\right)+2 b \left(b \ddot{a}+2 \dot{a} \dot{b}\right)}{a b^2},\\
    \mathcal{K}=\frac{8 \left(\frac{\dot{a}^2 \dot{b}^2}{a^2}+\ddot{b}^2\right)}{b^2}+\frac{4 \ddot{a}^2}{a^2}+\frac{\left(2 \dot{b}^2+k\right)^2}{b^4}.
\end{align}
The energy-momentum tensor/Einstein tensor components are given by
\begin{subequations}
\begin{align}
    8\pi T^t_{~t}=-\frac{ 2 \dot{b}^2+k}{2  b^2}-\frac{2 \dot{a} \dot{b}}{ a b},\label{rhoA}\\
    8\pi T^z_{~z}=-\frac{2\ddot{b}}{ b}-\frac{2 \dot{b}^2+k}{2 b^2},\label{pzA}\\
    8\pi T^r_{~r}=8\pi T^\theta_{~\theta}=-\frac{b \ddot{a}+a \ddot{b}+\dot{a} \dot{b}}{a b}\label{ptA}.
\end{align}
\label{TA}
\end{subequations}
From Eq. (\ref{TA}), metric (\ref{met}) generally describes a Segre [(11)1,1] spacetime with three independent functions specifying the Ricci sector. As a shorthand, we will write
\begin{align}
    -T^t_{~t}=\rho,\qquad T^z_{~z}=p_z,\qquad T^r_{~r}=T^\theta_{\theta}=p_\perp
\end{align}
Degeneration to [1(11,1)] occurs when $p_\perp=-\rho$, or in terms of the metric functions
\begin{align}
    2b(b \ddot{a}-\dot{a}\dot{b})=a(k+2\dot{b}^2-2b \ddot{b}) \label{degen}
\end{align}
This is slightly more complicated that the degeneration condition to [(11)(1,1)] (equation 2.9 in \cite{2023arXiv230109204B}) and a way to eliminate one of the functions in favor of the other is not obvious. However, when combined with an equation of state of the form
\begin{align}
    p_z=f(\rho)
\end{align} 
it is sometimes enough to find solutions.
The covariant conservation of energy equation $\nabla_\mu T^\mu_{~\nu}=0$, assuming $p_\perp=-\rho$ or equivalently  Eq. (\ref{degen}) has already been applied, the gives one nontrivial equation
\begin{align}
   -\frac{\dot{a}}{a}\big(\rho+p_z)=\dot{\rho}.\label{TConservation}
\end{align}
\section{Example: $p_z=0$}
\label{pz0}
One equation of state for which an exact solution is possible to find is 
\begin{equation}
    p_z=0
\end{equation}
because this equation of state results in a single differential equation for $b$ only, namely
\begin{align}
    k+2 \dot{b}^2+4 b \ddot{b}=0\label{beqpz0}.
\end{align}
Once a solution to Eq. (\ref{beqpz0}) is found, it can be used with Eq. (\ref{degen}) to find $a$ that results in the correct Segre type.
As it turns out, if we invert Eq. (\ref{beqpz0}) to find $t(b)$ rather than $b(t)$\footnote{A trivial solution $b=const$ exists for $k=0$, in which case inverting the function doesn't work. In this case, Eq. (\ref{degen}) dictates $a=A_1 t+A_0$. All the Riemann tensor components Eq.(\ref{ReimannA}) vanish, so this trivial solution corresponds to a version of Minkowski space.} the relationship is expressible in terms of elementary functions. We require the derivative replacement rules
\begin{subequations}
\begin{align}
  \overset{\therefore }{b}=-t'''\dot{b}^4+3\frac{ \ddot{b}^2}{\dot{b}},\\
  \ddot{b}=-t''\dot{b}^3,\\
  \dot{b}=\frac{1}{t'},
\end{align}
\label{Dinverter}
\end{subequations}
here primes denote derivatives with respect to $b$. With these substitutions, Eq. (\ref{beqpz0}) becomes
\begin{align}
    4b t''=t'(2+kt'^2).
\end{align}
This can be integrated once to obtain
\begin{align}
    t'=\pm \sqrt{\frac{2}{c/b-k}}.
\end{align}
The solution to the second integration can be written as 
\begin{subequations}
   \begin{align}
   t= Q\pm\frac{\sqrt{2} c \left(\sqrt{B} \sqrt{1-B}-\sin ^{-1}\left(\sqrt{B}\right)\right)}{\sqrt{k^3}}, \qquad k\neq0,c\neq0\label{tsol1a} \\
   t=Q\pm \sqrt{\frac{8b^3}{9c}},\qquad k=0,c\neq0\label{tsol1b}\\
   t=Q\pm\sqrt{\frac{2}{-k}}~b\qquad k<0,c=0.\label{tsol1c}
\end{align} 
\end{subequations}
where we introduce the abbreviation $B=\frac{b k}{c}$ in (\ref{tsol1a}).  In the $k\ne0,c\ne0$ case, $t$ is real if $B<0, k<0$ or $0\le B\le 1, k>0$.

Moving Eq. (\ref{degen}) into terms of $b$ derivatives rather than $t$ derivatives results in
\begin{align}
a(2t'+kt'^3+2b t'')=2b\big(b t' a''-a'(t'+bt'')\big)
\end{align}
Using Eqs. (\ref{tsol1a}-\ref{tsol1c}), we can obtain expressions for $a$ in terms of $b$
\begin{subequations}
\begin{align}
a=w_2 B^3 \sqrt{1-B}  \, _2F_1\left(\frac{3}{2},\frac{7}{2};\frac{9}{2};B\right)+\frac{w_1\sqrt{1-B} }{\sqrt{B}},\qquad k\neq 0,c\neq0 \label{asol1a}\\
a=\frac{v_1}{\sqrt{b}}+v_2 b^3,\qquad k=0,c\neq0\label{asol1b}\\
a=u_1 b^2+u_2\qquad k<0,c=0\label{asolc}
\end{align}
\end{subequations}
where $_2F_1$ is a Hypergeometric function. Notice that if we are in the $0\le B\le 1$ case, both branches of $a$ are real for real $w_1,w_2$. If we are in the $B<0$ case, we either require $w_1$ as 0 or as an imaginary constant for $a$ to be real.

\subsection{Curvature Quantities and Analysis}
In spite of the simplicity of the setup, the solutions break up into a lot of different cases, so it is conducive to cover the cases separately.
\subsubsection{$k\ne0$, $c\neq0$}
Recall in this case we have 
\begin{align}
  t= Q\pm\frac{\sqrt{2} c \left(\sqrt{B} \sqrt{1-B}-\sin ^{-1}\left(\sqrt{B}\right)\right)}{\sqrt{k^3}},\quad
a=w_2 B^3 \sqrt{1-B}  \, _2F_1\left(\frac{3}{2},\frac{7}{2};\frac{9}{2};B\right)+\frac{w_1\sqrt{1-B} }{\sqrt{B}}.
\end{align}

The energy momentum tensor is given by
\begin{align}
T^t_{~t}=T^r_{~r}=T^\theta_{~\theta}=\frac{-7k^3 w_2}{16 \pi  c^2 a}
\end{align} 
with all other components being 0. This behavior agrees with what we would expect from the energy conservation equation, namely $\rho\propto1/a$. Since $w_2$ is in the numerator, if $a$ contains only the $w_1$ branch then there is no energy-momentum content. Naturally the Ricci scalar takes a similar form to the Einstein tensor component
\begin{align}
    \mathcal{R}=\frac{21 k^3 w_2}{2 c^2 a}.
\end{align}
The $Q_{11}$ matrix component which describes the Weyl behavior is also very simple
\begin{align}
Q_{11}=\frac{c}{2 b^3}.
\end{align}
The orthonormal Reimann components can be written
\begin{subequations}
\begin{align}
    R_{\hat{0}\hat{1}\hat{0}\hat{1}}=\frac{-7k^3 w_2}{4 c^2 a}-\frac{c}{2b^3},\\
    R_{\hat{0}\hat{2}\hat{0}\hat{2}}=R_{\hat{0}\hat{3}\hat{0}\hat{3}}=\frac{c}{4b^3}\\
    R_{\hat{2}\hat{3}\hat{2}\hat{3}}=\frac{c}{2b^3}\\
     R_{\hat{3}\hat{1}\hat{3}\hat{1}}= R_{\hat{1}\hat{2}\hat{1}\hat{2}}=\frac{7k^3 w_2}{4 c^2 a}-\frac{c}{4b^3}.
\end{align}
\label{Reimannpz0}
\end{subequations}
The Kretchman scalar can be written as 
\begin{align}
\mathcal{K}=4\Big(\big(\frac{-7k^3 w_2}{4 c^2 a}-\frac{c}{2b^3}\big)^2+6\big(\frac{c}{4b^3}\big)^2+2\big(\frac{7k^3 w_2}{4 c^2 a}-\frac{c}{4b^3}\big)^2\Big)
\end{align}

Notice that when the curvature functions are written in terms of $b$, the constant $Q$ and the sign from Eq. (\ref{tsol1a})  do not directly appear in the curvature quantities. Also, all the above curvature quantities are expressed in terms of the two items
\begin{align}
    \frac{k^3w_2}{c^2 a},\qquad \frac{c}{b^3}
\end{align} 
which are the Ricci scalar and Weyl $Q_{11}$ term up to constants. These are singular if $a$ or $b$ go to 0. 

\begin{figure}
    \centering
    \includegraphics[width=7cm]{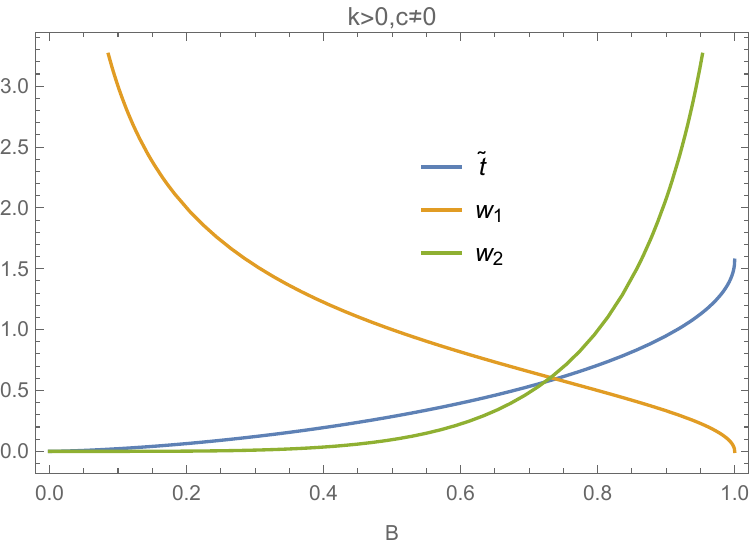}
    \includegraphics[width=7cm]{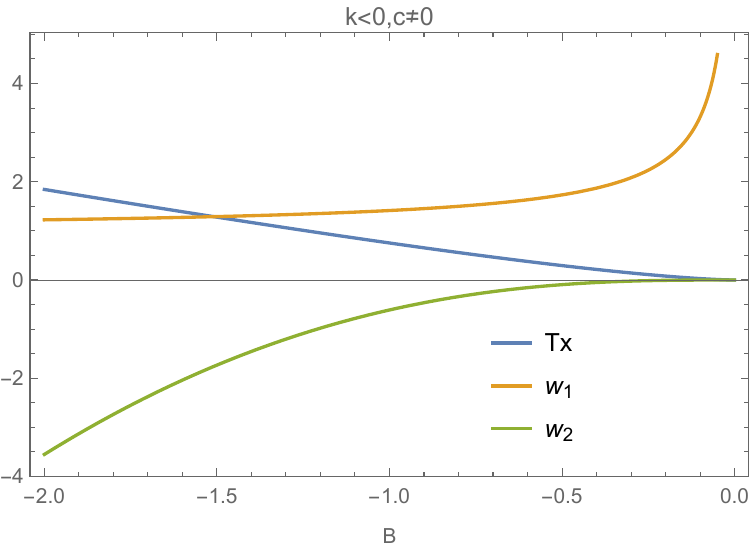}
    \caption{Plots of the behavior of $a$ and $t$ when $k\ne0$, $c\ne0$. We have two separate cases, one with positive $k$ where $0<B<1$ and one with negative $k$ where $B<0$. }
    \label{pz0A}
\end{figure}

If $k>0$, $0\le B \le 1$ for $t'=\sqrt{\frac{2}{k/B-k}}$ to be real and $0\le\bar{t}\le \pi/2$ so the solution must be bounded on two sides. For $k<0$, $B\le 0$ and $\bar{t}\ge0$ so we must only one side be bounded on one side. Additional singularities may show up for particular values of $w_1$, $w_2$ if they are such that $a=0$ at some time. We show plots for the behavior of $a$ and $t$ in terms of $b$ in Figure \ref{pz0A} for $k>0, c\ne0$, $0\le B\le 1$. We show  $\tilde{t}=\left(\sin ^{-1}\left(\sqrt{B}\right)-\sqrt{B} \sqrt{1-B}\right)=\mp (t-Q)\sqrt{k^3/2}/ c$, $a(w_1=1,w_2=0)$ and $a(w_2=1,w_1=0)$. For $k<0,c\ne0$, $B<0$ we show $T_x=i\left(\sin ^{-1}\left(\sqrt{B}\right)-\sqrt{B} \sqrt{1-B}\right)=\mp (t-Q)\sqrt{|k|^3/2}/ c$, $a(w_1=i,w_2=0)$ and $a(w_2=1,w_1=0)$.

\subsubsection{$k=0$, $c\neq0$}

\begin{figure}[h]
    \centering
    \includegraphics[width=7cm]{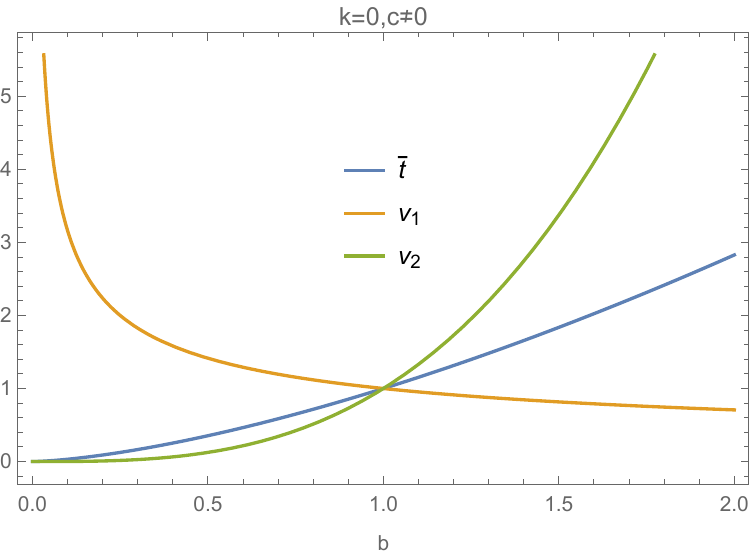}
    \caption{Plots of the behavior of $a$ and $t$ for $k=0,c\ne0$. The label $v_1$ corresponds to $a(v_1=1,v_2=0)$, $v_2$ corresponds to $a(v_1=0,v_2=1)$. There is a singularity at $b=0$, $\bar{t}=0$ in the Weyl sector which could act as a boundary of this universe. }
    \label{pz0B}
\end{figure}

In this case, we have
\begin{align}
     t=Q\pm \sqrt{\frac{8b^3}{9c}}, \qquad a=\frac{v_1}{\sqrt{b}}+v_2 b^3.
\end{align}
Since $t(b)$ is simpler in the $k=0$ case Eq. (\ref{tsol1b}) it is possible to invert it, giving 
\begin{align}
b=\Bigg(\frac{9c(t-Q)^2}{8}\Bigg)^{1/3}
\end{align}
 where it is assumed that $b$ and $c$ are positive. It is therefore possible to rewrite all of these items as explicit functions of $t$, but doing so is typically less compact than writing them in terms of $a,b$.
Overall the curvature quantities follow the same basic structure as in the generic case with nonzero $k$. The energy momentum tensor is given by
\begin{align}
T^t_{~t}=T^r_{~r}=T^\theta_{~\theta}=\frac{-7 c ~v_2}{16 \pi a}
\end{align} 
with all other components being 0. Again we have $\rho\propto1/a$, as well as the fraction $-7/16$. The Ricci scalar is
\begin{align}
    \mathcal{R}=\frac{21 c ~v2}{2a}
\end{align}
The $Q_{11}$ matrix component is the same as in the $k\ne0$ case
\begin{align}
Q_{11}=\frac{c}{2b^3}
\end{align}
The orthonormal Reimann components can be written
\begin{subequations}
\begin{align}
    R_{\hat{0}\hat{1}\hat{0}\hat{1}}=-\frac{7 c ~v_2}{4a}-\frac{c}{2b^3},\\
    R_{\hat{0}\hat{2}\hat{0}\hat{2}}=R_{\hat{0}\hat{3}\hat{0}\hat{3}}=\frac{c}{4b^3}\\
    R_{\hat{2}\hat{3}\hat{2}\hat{3}}=\frac{c}{2b^3}\\
     R_{\hat{3}\hat{1}\hat{3}\hat{1}}= R_{\hat{1}\hat{2}\hat{1}\hat{2}}=\frac{7 c ~v_2}{4a}-\frac{c}{4b^3}
\end{align}
\label{Reimannpz0}
\end{subequations}
The Kretchman scalar can be written
\begin{align}
\mathcal{K}=4\Big( \big(-\frac{7 c ~v_2}{4a}-\frac{c}{2b^3}\big)^2+6\big(\frac{c}{4b^3}\big)^2+\big(\frac{7 c ~v_2}{4a}-\frac{c}{4b^3}\big)^2\Big)
\end{align}
Once again we have two terms from which the curvature quantities are built, being basically the Ricci scalar and $Q_{11}$
\begin{align}
   \frac{c ~v_2}{a},\qquad \frac{c}{b^3}
\end{align}
or as explicit functions of time
\begin{align}
\frac{c v_2}{\frac{\sqrt{2} v_1}{3^{1/3}(c(t-Q)^2)^{1/6}}+\frac{9 c v_2}{8}(t-Q)^2},\quad \frac{8}{9(t-Q)^2}
\end{align}
For $k=0,c\ne0$, we show $\Bar{t}=t^{3/2}=\pm 3(t-Q)\sqrt{c/8}$, $a(v_1=1,v_2=0)$, and $a(v_2=1,v_1=0)$ in Figure \ref{pz0B}. This universe is bounded by $b=0,\bar{t}=0$, at which point a singularity is present in the Weyl sector. There may be additional singularities in the Ricci sector if $a=0$, depending on the values of the $v_1$ and $v_2$ constants, that could possibly act as boundaries.
\subsubsection{$k<0$,~$c=0$}
Recall that now
\begin{align}
    t=Q\pm\sqrt{\frac{2}{-k}}~b \qquad a=u_1 b^2+u_2.
\end{align}
Notice that in this case we have a simple linear relationship between $t$ and $b$, that can be trivially inverted to give
\begin{align}
    b=\pm\sqrt{\frac{-k}{2}}(t-Q),
\end{align}
The Energy tensor components are
\begin{align}
T^t_{~t}=T^r_{~r}=T^\theta_{~\theta}=\frac{k u_1}{4\pi  a}.
\end{align} 
The Weyl behavior vanishes in this case, we obtain 
\begin{align}
    Q_{11}=0
\end{align}
Because the Weyl behavior vanishes with the constant $c$, the orthonormal Reimann components simplify such that
\begin{align}
   R_{\hat{0}\hat{1}\hat{0}\hat{1}}= -R_{\hat{3}\hat{1}\hat{3}\hat{1}}= -R_{\hat{1}\hat{2}\hat{1}\hat{2}}=\frac{k u_1}{a}
\end{align}
with nonlisted components being identically 0.
The Ricci and Kretchman scalars are 
\begin{align}
    \mathcal{R}=\frac{-6k u_1}{a},~\mathcal{K}=\frac{12 k^2 u_1^2}{a^2}
\end{align}
With the vanishing of the Weyl behavior, we effectively only have one curvature parameter 
\begin{align}
   \frac{u_1 k}{a}= \frac{k u_1}{u_1 \frac{-k}{2}(t-Q)^2+u_2}.
\end{align}
 In the case where $u_1\neq0$, this situation describes curvature/ energy quantities diluting as $a$. If $u_1=0$, all curvature quantities are 0, so this is an unusual coordinate system of Minkowski space. It is possible for their to be singularities if $a=0$ for particular $u_1,u_2$, but it is also possible for this universe to have no singularities bounding it in either direction because the $b=0$ singularity present in the Weyl sector for $c\ne0$ has vanished with $c$.
For $k<0,c=0$, we show $\hat{t}=\pm(t-Q)\sqrt{-k/2}$, as well as $a(u_1=1,u_2=0)$ and $a(u_1=0,u_2=1)$. 

\begin{figure}[h]
    \centering
     \includegraphics[width=7cm]{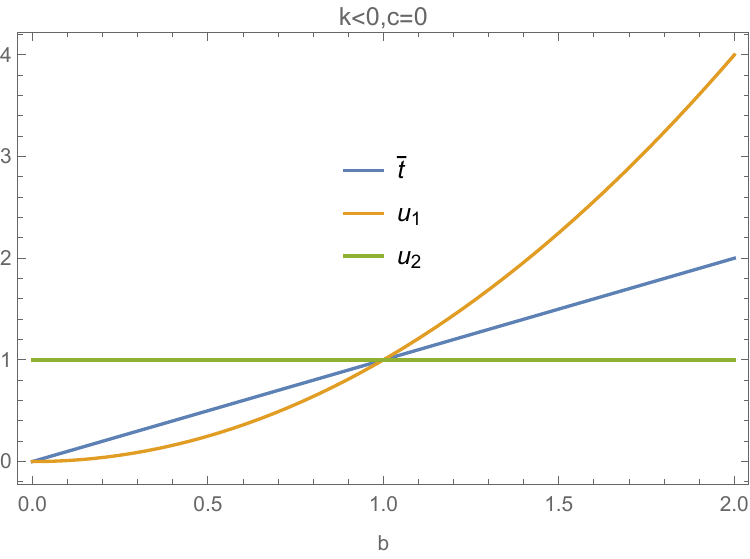}
    \caption{Plots of the behavior of $a$ and $t$ for $k<0,c=0$. The label $u_1$ corresponds to $a(u_1=1,u_2=0)$, the label $u_2$ corresponds to $a(u_1=0,u_2=1)$. In this case, there are no restrictions on $b$ imposed by $t$ being real.}
    \label{pz0C}
\end{figure}

 \section{Example: $p_z=-\rho$}
 The case where $p_z=-\rho$ for Segre type [1(11,1)] is ultimately the same as the case $p_\perp=-\rho$ for Segre type [(11)(1,1)]. They both degenerate to vacuum energy Segre type [(111,1)], and for the starting metric (\ref{met}) result in the same set of differential equations, being either
 \begin{align}
 a=&c\dot{b},&&k=\frac{2b^2 \overset{\therefore}{b}}{\dot{b}}-2\dot{b}^2 \qquad \text{or}\label{vacA}\\
 b=&B,&&\frac{\ddot{a}}{a}=\frac{k}{2B^2}.\label{vacB}
 \end{align}
The behavior for these cases was previously explored in \cite{2023arXiv230109204B}. 

Specifically, Eq.~(\ref{vacB}) results in a spacetime for which both the Weyl and Ricci sectors of the curvature are proportional to $k/B^2$ and do not evolve with time, as opposed to the vanishing Weyl tensor of de Sitter space and the time varying Weyl tensor from the solution to Eq.~(\ref{vacA}) when $W\ne0$. In the case $k=0$, the Riemann tensor components all vanish indicating the solution degenerates to Minkowski space. The other scale factor function $a$ evolves as
\begin{align}
   a=X \cosh(\sqrt{\frac{k}{2B^2}}t)+Y \sinh(\sqrt{\frac{k}{2B^2}}t) 
\end{align}
This solution is explored in more detail in Appendix C of \cite{2023arXiv230109204B}.

Conversely, Eq.~(\ref{vacA}) results in a vacuum energy spacetime with a potentially nontrivial Weyl sector as opposed to the identically vanishing Weyl tensor of standard De Sitter space. It is possible in any case to parameterize this solution with a condition 
\begin{align}
    t'=\pm\sqrt{\frac{6b}{6 b^3 V-3 b k-4 W}}\propto 1/a.\label{avack}
\end{align}

In this case the orthonormal Riemann components are
\begin{align}     
      R_{\hat{0}\hat{1}\hat{0}\hat{1}}= -R_{\hat{2}\hat{3}\hat{2}\hat{3}}=\frac{2W}{3b^3}-V,\\
      R_{\hat{0}\hat{2}\hat{0}\hat{2}}=R_{\hat{0}\hat{3}\hat{0}\hat{3}}=-R_{\hat{3}\hat{1}\hat{3}\hat{1}}=-R_{\hat{1}\hat{2}\hat{1}\hat{2}}=-\frac{W}{3b^3}-V.
\end{align}
The Ricci scalar is $12 V$ and $Q_{11}=\frac{-2W}{3b^3}$, and all the other curvature quantities can be expressed in $V$ and $W/b^3$ type terms. Interestingly, we have $Q_{11}\propto b^{-3}$ in both this case and the $p_z=0$ case, so there are curvature singularities when $b=0$ unless $W=0$. We have our energy/momentum tensor components going like constants instead of $1/a$ because Eq. (\ref{TConservation}) goes to $\dot{\rho}=0$ for this equation of state.
For general $k,V,W$, it is not possible to find an expression for $b(t)$ or $t(b)$, but there are some cases where it can be done.

In the $k=0$ case it is simple enough to invert the expressions and obtain explicit functions of $t$
\begin{align}
   b=X \sqrt[3]{e^{-2 Y (t+Z)}+e^{2 Y (t+Z)}-2},\qquad a=c\frac{2 X Y \left(e^{2 Y (t+Z)}-e^{-2 Y (t+Z)}\right)}{3 \big(e^{-2 Y
   (t+Z)}+e^{2 Y (t+Z)}-2\big)^{2/3}},\label{bvac0}
\end{align}
where $ V=\frac{4Y^2}{9},~W=\frac{-8X^3 Y^2}{3}$. More specific information about this, such as other curvature quantities graphs for the time dependence functions are given in 2.3 of \cite{2023arXiv230109204B}.

It was not mentioned in \cite{2023arXiv230109204B}, but it is also possible to solve Eq.~(\ref{avack}) in the $W=0$ and $V=0$ cases. 
\subsection{$W=0$ }
For $W=0$,  we obtain 
\begin{align}
   t'=\pm\sqrt{\frac{2}{2 b^2 V-k}},
\end{align}
from which we see that is real a) everywhere if $k<0,V>0$, b) nowhere if $k>0,V<0$, c) $b^2\ge\frac{k}{2V}$ if $V>0,k>0$, d) $b^2\le\frac{k}{2V}$ if $k<0,V<0$. In the cases a,c,d where $t'$ is real, we obtain
\begin{align}
    t+Q=\pm\frac{\tanh^{-1}(b\sqrt{\frac{2V}{2b^2 V-k}})}{\sqrt{V}}.\label{tw0}
\end{align}
In case c), the constant $Q$ has to have an imaginary component $i\pi/2$ for $b$ and $t$ to be real, but in the a) and d) cases $Q$ should be real.

\begin{figure}[h]
    \centering
    \includegraphics{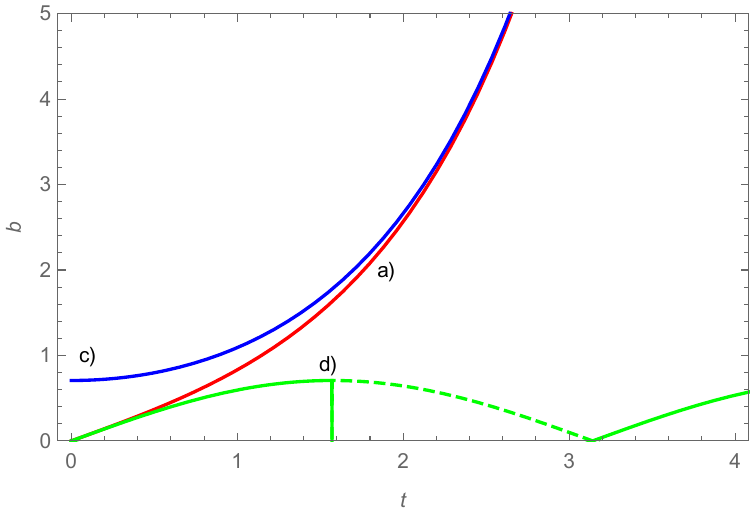}
    \caption{Basic behavior of $t$ vs $b$ for the $W=0$ cases a) (specifically plotting $k=-1,V=1,Q=0$ ), c) (specifically plotting $k=V=1,Q=i\pi /2$ ), and d) (specifically plotting $k=V=-1,Q=0$ ). Notice that in case d), the $+$ branch is shown in solid and the $-$ branch is dashed, because Eq \ref{btw0} has sign changes as written. However, everything in the metric and surviving curvature terms have even powers of $b$ in this spacetime so the sign is not really relevant here, the scale $b^2$ in the metric  seems to oscillate in case d). In cases a), c), we have similar expansion as $t$ increases.}
    \label{pzm}
\end{figure}

It is possible to recover an expression for $b(t)$ from Eq. (\ref{tw0}), specifically
\begin{align}
b(t)=\pm \frac{\sqrt{k} \tanh \left((t+Q) \sqrt{V}\right)}{\sqrt{V} \sqrt{2 \tanh
   ^2\left((t+Q) \sqrt{V}\right)-2}}\, .\label{btw0}
\end{align}
We show the behavior of this in the Figure \ref{pzm}. The cases a) and c)  both feature similar expansion at large $t$ and are the positive $V$. The d) case with negative $V$ is possibly oscillatory in scale factor.

Notice that with $W=0$, the Weyl behavior drops out and we are left with only the constant Ricci curvature in terms of $V$. This might therefore be de Sitter space (or anti de Sitter for $V<0$) in unusual coordinates.
\subsection{$V=0$ }
\begin{figure}[h]
    \centering
    \includegraphics{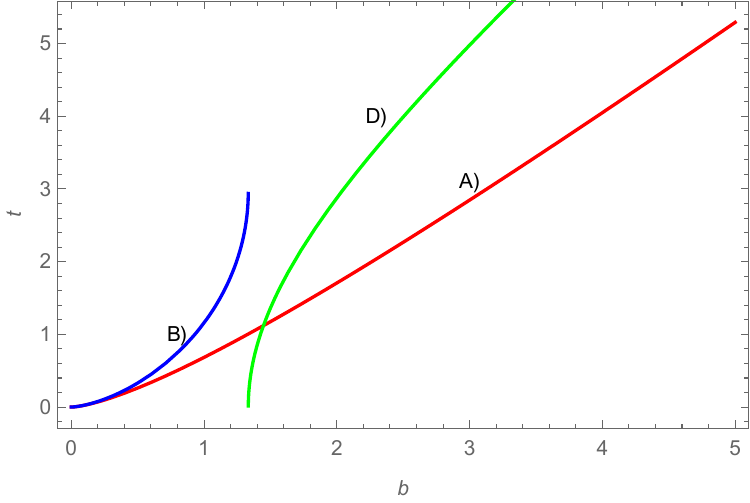}
    \caption{Graphs of cases A) (specifically $k=-1,W=-1, Q=0$), B) (specifically $k=1,W=-1,Q=0$), and D) (specifically $k=-1,W=1,Q=\frac{2i\sqrt{2}\pi}{3}$. Notice that case B) only is real for $b<\frac{4W}{-3k}=4/3$ and case D) only is real for $b>\frac{4W}{-3k}=4/3$. Interestingly, the $t$ value at which branch B stops being fully real is $\frac{2\sqrt{2}\pi}{3}$, which is the same magnitude as the required imaginary component in case D).  }
    \label{pzmv0}
\end{figure}

In the $V=0$ case, we have
\begin{align}
    t'=\pm\sqrt{\frac{6b}{-3 b k-4 W}}\label{tv0}
\end{align}
which is real for positive $b$ A) everywhere if $k<0,W<0$, B) $b\le\frac{4W}{-3k}$ if $k>0,W<0$, C) nowhere if $k>0,W>0$, D) $b\ge\frac{4W}{-3k}$ if $k<0,W>0$. It is possible to integrate Eq. (\ref{tv0}) in the A), B), D) cases
\begin{subequations}
\begin{align}
   t+Q&=\pm\frac{\sqrt{2}  \left(\sqrt{3 b k (3 b k+4 W)}+4 W \sinh
   ^{-1}\left(\sqrt{\frac{3 b k}{4 W}}\right)\right)}{3 k \sqrt{-k}},\text{ A), B)}\label{tvAB}\\
   t+Q&=\pm\frac{\sqrt{2}\left(\sqrt{-3 b k (3 b k+4 W)}+4 W \sin
   ^{-1}\left(\sqrt{-\frac{3 b k}{4 W}}\right)\right)}{3 k^{3/2}},\text{ D).}\label{tvD}
\end{align}
\end{subequations}
Notice that the form of the equations is the same in the $W<0$ cases A), B) (Eq.\ref{tvAB}), and that in the $W>0$ case D) (Eq.\ref{tvD}) we need to have $Q$ have an imaginary component of $\frac{2i\sqrt{2}\pi}{3}$ in order for $t$ to be real. There does not seem to be an obvious way to invert these expressions to find $b(t)$, but in the cases A) and D) where it is unbounded $b$ is unbounded from above, they approach a linear relation $t\approx\pm b\sqrt{2/-k} $ as $b$ increases. We plot the behavior in Figure \ref{pzmv0}.

Since $V=0$ sends the components of $T^\mu_{~\nu}$ to zero, this is a pure vacuum case with a nontrivial Weyl behavior.

\section{Example: $p_z=\rho$}
Usage of the equation of state $p_z=\rho$ results in the condition
\begin{align}
\frac{\dot{a}}{a}=-\frac{k+2 \dot{b}^2+2 b \ddot{b}}{2 b \dot{b}}\label{azrho}
\end{align}
assuming a nonconstant $b$. This ultimately allows for the elimination of $a$ from Eq. (\ref{degen}), but results in a third order differential equation for $b$, namely
\begin{align}
 4b^2 \dot{b}~\overset{\therefore}{b}=8 \dot{b}^4+(6\dot{b}^2+k+4 b\ddot{b})(k+2 b \ddot{b})
 \label{bzrho}
\end{align} 
for which we did not find a solution.

It is possible in any case to use conservation of energy. Using the equation of state and Eq.~(\ref{TConservation}) we get
\begin{align}
\rho=\frac{Q}{a^2}
\end{align}
where $Q$ is an undetermined constant.

\subsection{Some Numerical Solutions of \ref{bzrho}}
If one has initial conditions for $b,b',b''$, and a value for $k$, it is possible to numerically solve Eq.(\ref{bzrho}), then use the numerical solution and Eq.(\ref{azrho}) to find
\begin{subequations}
\begin{align}
a&=C e^{\int-\frac{k+2 \dot{b}^2+2 b \ddot{b}}{2 b \dot{b}} dt}\label{anumA}\\
a&=\frac{Ce^{-\int\frac{k}{2 b \dot{b}} dt}}{\dot{b} b}\label{anumB}
\end{align}
\end{subequations}
where $C$ is a constant. However, since the curvature components are all in terms of $\ddot{a}/a$ or $\dot{a}/a$ (see Eqs.\ref{ReimannA}-\ref{TA}), the $C$ constant is a gauge parameter and can be set to 1 for convenience. Once numerical solutions for $a$ and $b$ exist, the relevant curvature parameters can be computed. 
In these examples, there are three error parameters that we  track to determine whether the solution behaves well. The term
\begin{subequations}
\label{errordefs}
\begin{align}
    Er_{Segre}=p_\perp+\rho,
\end{align}
where $p_\perp$ and $\rho$ are calculated using the expressions in Eq. (\ref{TA}), should be identically 0 for a perfect solution because of the [1(11,1)] Segre type. Likewise, the term
\begin{align}
Er_{EOS}=p_z-\rho
\end{align}
would be identically 0 for a perfect solution because of the supposed $p_z=\rho$ equation of state. Finally, the energy conservation equation means that 
\begin{align}
    Er_{Cons}=a(t)^2\rho(t)-a(0)^2\rho(0)^2
\end{align}
\end{subequations}
should also be identically 0 for a perfect solution. These error parameters are computed by creating interpolating functions in Mathematica \cite{Mathematica} for $a$ and $b$.

In order to examine some of the possibilities of these solutions, we track the evolution of the scale factors, error parameters, Energy-Momentum tensor eigenvalue $p_z$ (note that $p_\perp$ and $\rho$ are trivially related for this system), and Kretchman scalar.
\subsubsection{k=0}
The $k=0$ case is substantially easier from a computational standpoint because no numerical integration needs to be done to solve for $a$. We can find an interpolating function for $b$ from Eq. (\ref{bzrho}) and then we can simply use $a=1/(b\dot{b})$, as the integral in the exponent of Eq.(\ref{anumB}) is 0.
We examine the initial condition $b(0)=1$, $\dot{b}(0)=\pm1$, and $\ddot{b}(0)=(-1,0,1)$ \footnote{we don't directly consider cases with $b=0$ or $\dot{b}=0$ for the initial conditions because this leads to problems for $a$} , which leads to 6 cases, and evolve the state forward and backward from $t=0$. However, since the cases with $\dot{b}=-1$ are equivalent to the time reverse of the corresponding cases with $\dot{b}=1$ (other than an irrelevant sign for $a$), we only plot the results from the $\dot{b}=1$.

\begin{figure}[h]
    \centering
    \includegraphics[width=7cm]{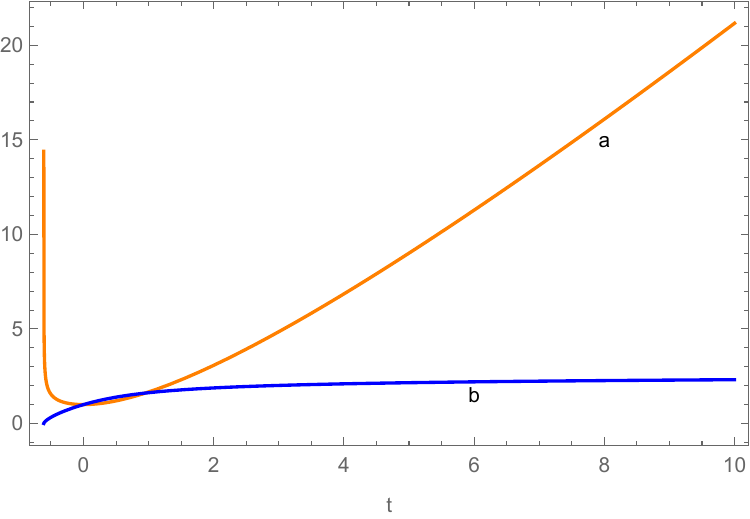}
    \includegraphics[width=7cm]{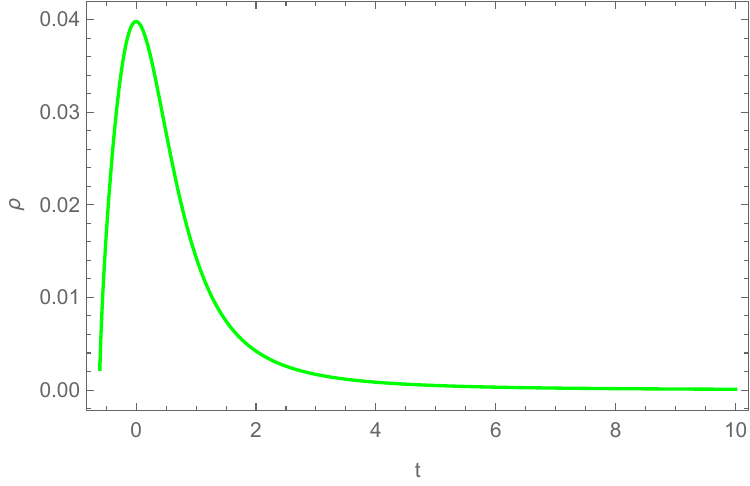}
    \includegraphics[width=7cm]{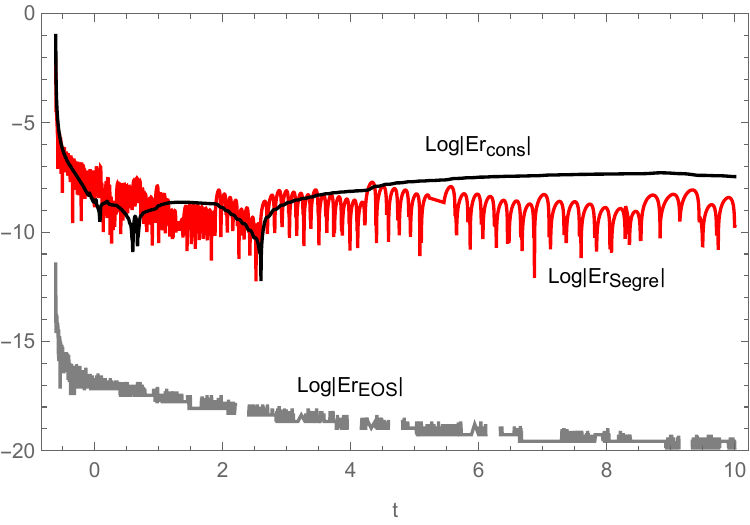}
    \includegraphics[width=7cm]{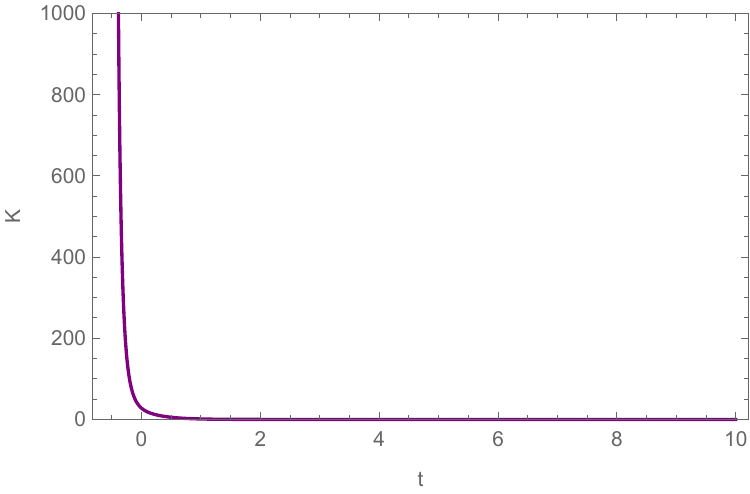}
    \caption{Plots of the scale functions $a,b$ (top left panel), $\rho=p_z=-p_\perp$ (top right panel), Error parameters \ref{errordefs} (bottom left panel), and Kretchman scalar (bottom right panel) for $k=0,\dot{b}=1,\ddot{b}=-1$. Note the singularity at $t\approx-0.609$, at which point $b\rightarrow 0$, the Kretchman scalar appears to diverge, and the error parameters are at their largest. The $Er_{EOS}$ is usually smaller in magnitude than $Er_{cons}$ or $Er_{Segre}$, except at locations where the error changes in sign (visible for instance as the cusps in $Er_{cons}$ at $t\approx 0.5$, $t\approx 2.5$.}
    \label{fig:kz1}
\end{figure}

For $\dot{b}=1$, $\ddot{b}=-1$, we have a singularity at $t\approx -0.609$ where the scale factor $b$ goes to $0$ and the Kretchman scalar appears to diverge, although the energy momentum tensor eigenvalue seems to go to 0. At large positive $t$, we appear to approach a configuration where the Kretchman scalar and energy momentum tensor eigenvalue goes to 0, $a$ expands roughly linearly with time, and $b$ expands extremely slowly. The energy momentum tensor eigenvalue peaks at around $t=0$. We plot these functions in Figure \ref{fig:kz1}, as well as plots of the log of the absolute value of the error parameters.

For $\dot{b}=1$, $\ddot{b}=0$, we have a solution with singularities at $t\approx \pm 0.73$. At the negative time singularity, $b$ goes to 0, and at the positive time singularity $a$ goes to 0. Both singularities have a divergence in the Kretchman scalar, but only the positive time singularity has a divergence in the density. This solution has negative density and therefore violates the WEC.

\begin{figure}[h]
    \centering
    \includegraphics[width=7cm]{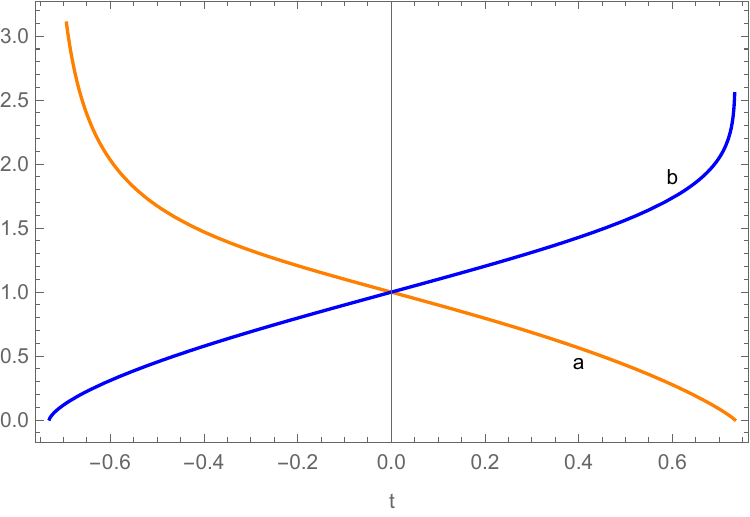}
    \includegraphics[width=7cm]{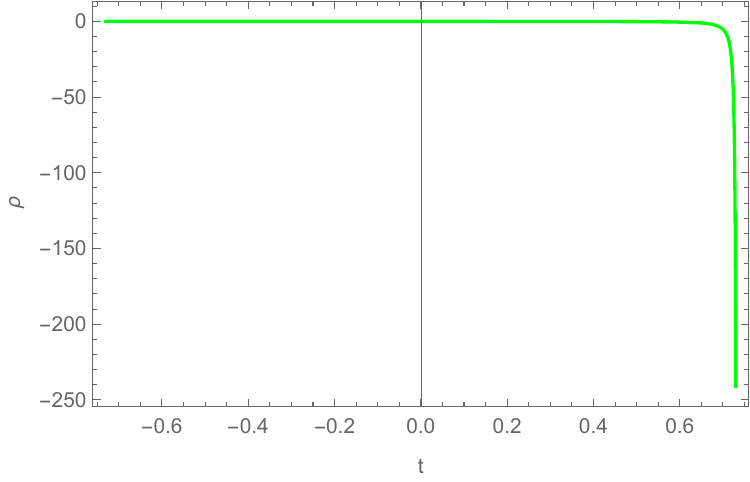}
    \includegraphics[width=7cm]{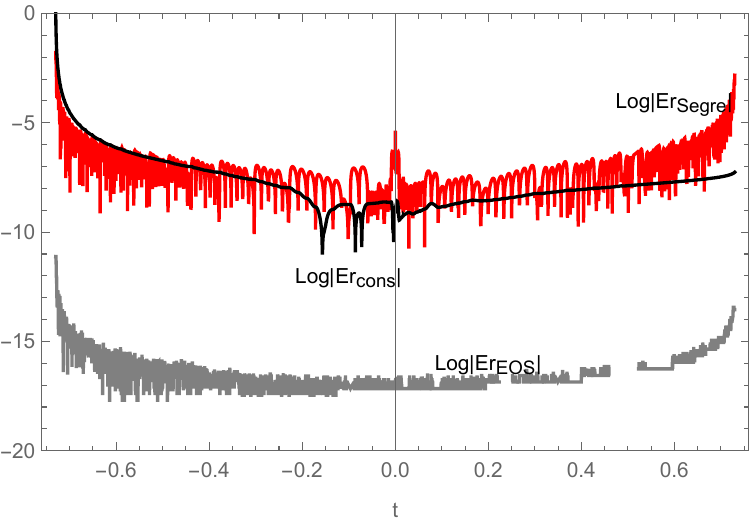}
    \includegraphics[width=7cm]{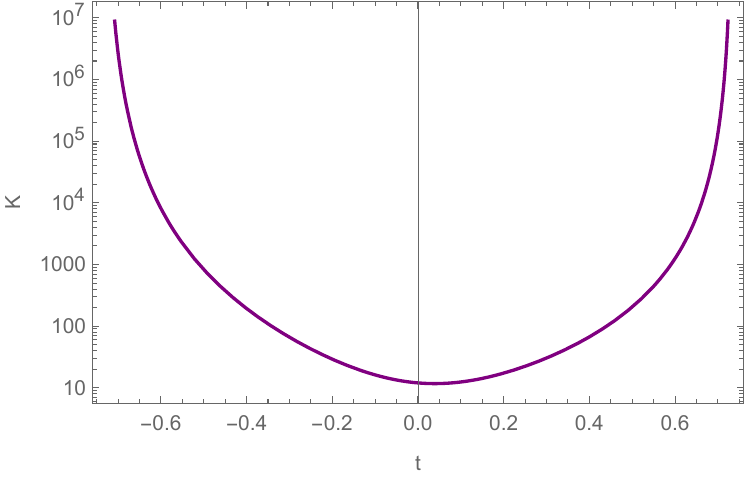}
    \caption{Plots of the scale functions $a,b$ (top left panel), $\rho=p_z=-p_\perp$ (top right panel), Error parameters \ref{errordefs} (bottom left panel), and Kretchman scalar (bottom right panel) for $k=0,\dot{b}=1,\ddot{b}=0$. We are bounded on both sides by singularities at $t\approx\pm0.73$ in this case. Once again $Er_{EOS}$ is usually smaller in magnitude than $Er_{cons}$ or $Er_{Segre}$, but all errors have a noticeable increase as we approach the singularities.}
    \label{fig:kz2}
\end{figure}

 The case $\dot{b}=1$, $\ddot{b}=1$, is qualitatively very similar to the behavior of the $\dot{b}=1$, $\ddot{b}=0$ case, except the singularities are now located at $t\approx-0.88$ and $t\approx0.365$.

\begin{figure}[h]
    \centering
    \includegraphics[width=7cm]{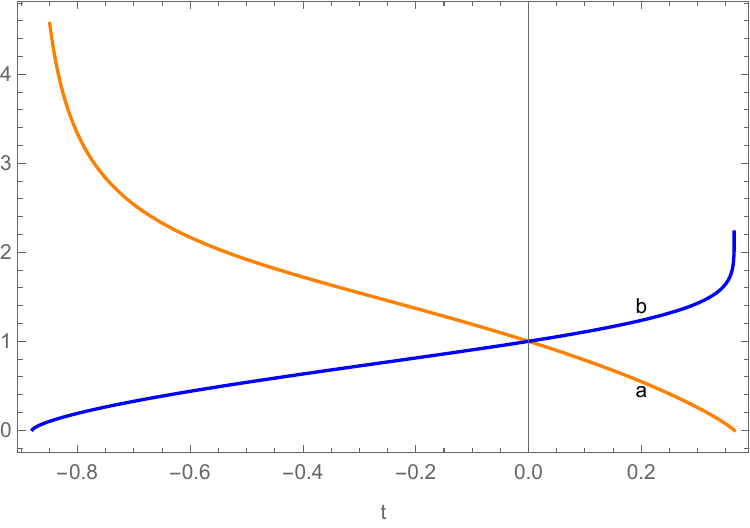}
    \includegraphics[width=7cm]{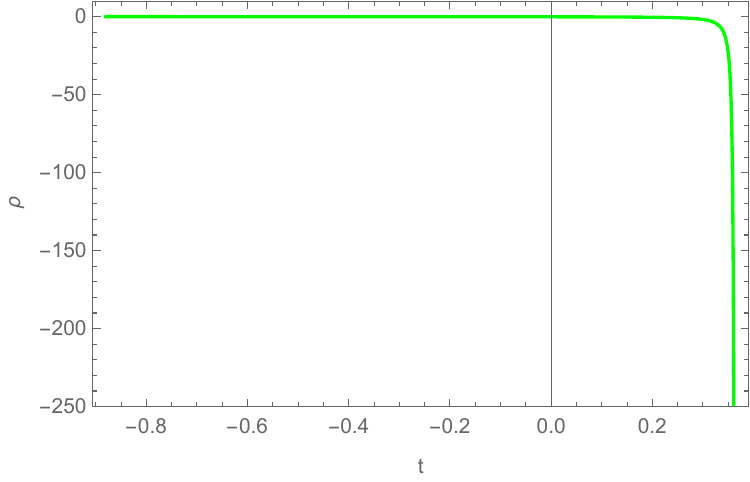}
    \includegraphics[width=7cm]{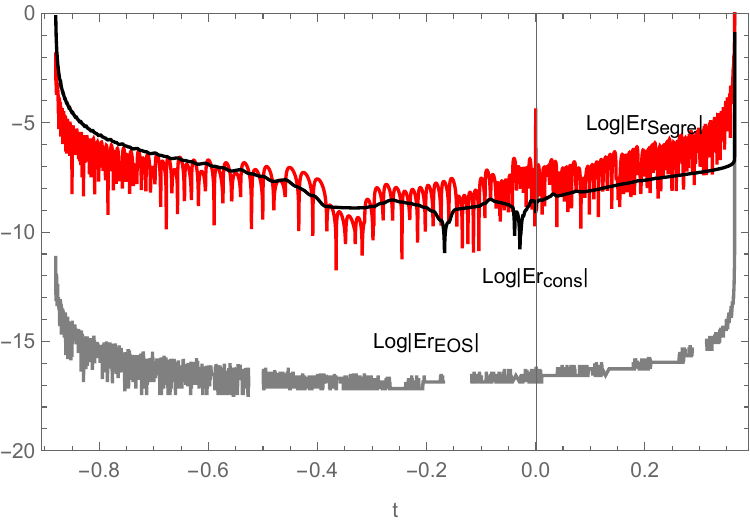}
    \includegraphics[width=7cm]{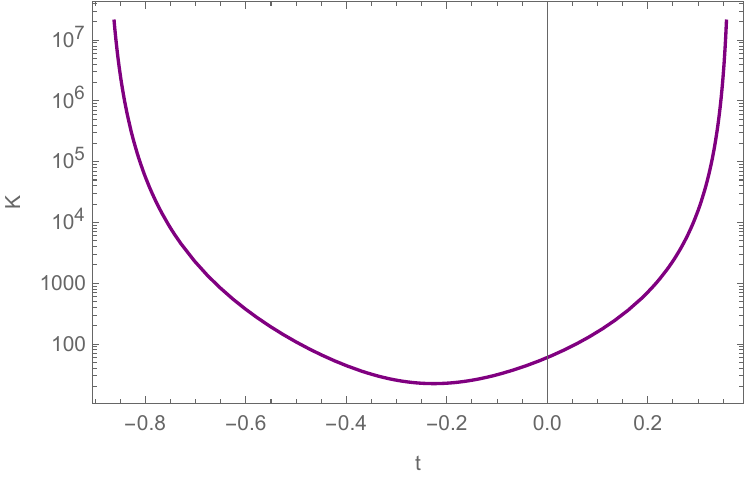}
    \caption{Plots of the scale functions $a,b$ (top left panel), $\rho=p_z=-p_\perp$ (top right panel), Error parameters \ref{errordefs} (bottom left panel), and Kretchman scalar (bottom right panel) for $k=0,\dot{b}=1,\ddot{b}=1$. We are bounded on both sides by singularities at at $t\approx-0.88$ and $t\approx0.365$. The behavior is otherwise qualitatively similar to the $\ddot{b}=0$ case.}
    \label{fig:kz3}
\end{figure}
\subsubsection{k=1}
While it is possible in principle to use Eq. (\ref{bzrho}) to find $b$ and then use Eq. (\ref{anumA}) or Eq. (\ref{anumB}) to find $a$, the implementation of doing this leads to very slow performance due to the integration step. Further, Eq. (\ref{anumA}) lead to unacceptably large error parameters. However, it is possible to compute interpolating functions for $a$ and $b$ simultaneously by solving the coupled equations
\begin{subequations}
    \begin{align}
    2b \dot{a} \dot{b}=-a((k+2\dot{b}^2)+2 b \ddot{b}),\\
    2b\dot{a} \dot{b}+a(k+2\dot{b}^2)=2b(b\ddot{a}+a\ddot{b})
\end{align}
\label{abcoupled}
\end{subequations}
which are equivalent to the equation of state constraint $p_z=\rho$ and the Segre constraint $p_\perp=-\rho$. Using Eq. (\ref{abcoupled}) allows for accurate interpolating functions to be computed in a reasonable time.

It is again the case that switching the sign of the initial condition of $\dot{b}$ basically creates a time reverse solution, which agrees with the structure of the equations in Eq. (\ref{abcoupled}). Qualitatively, the $k=1,\ddot{b}=0,1$ cases have a lot in common with the $k=0,\ddot{b}=0,1$ cases we examined in the previous subsection. They are all bounded by singularities on both sides, have one scale function going to 0 at the initial singularity and the other at the final singularity, and feature negative diverging energy density. The $k=1,\dot{b}=-1$ case differs from $k=0,\dot{b}=-1$ because for $k=1$ we have the $b$ scale function going back to 0 and another singularity after a finite time.

\begin{figure}[h]
    \centering
    \includegraphics[width=7cm]{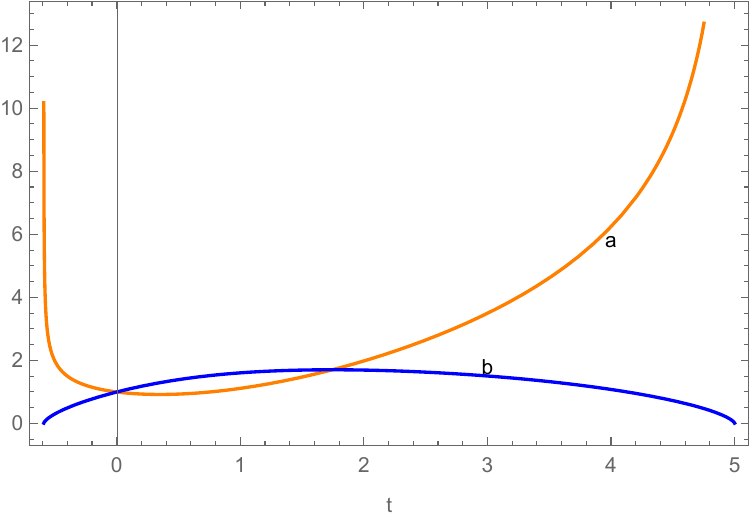}
    \includegraphics[width=7cm]{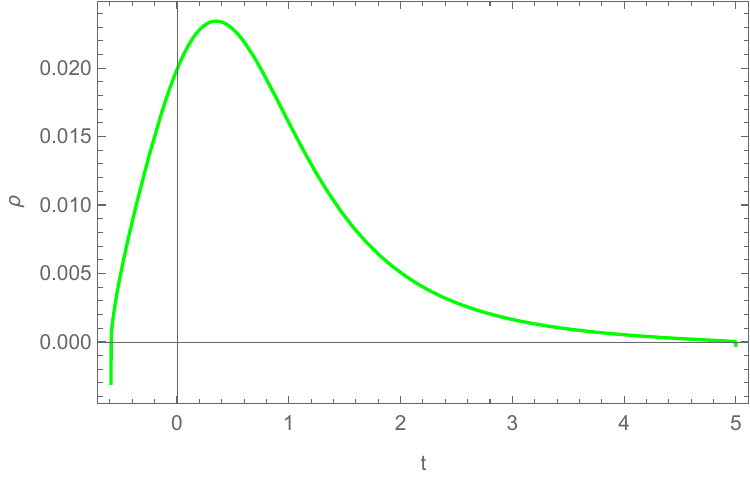}
    \includegraphics[width=7cm]{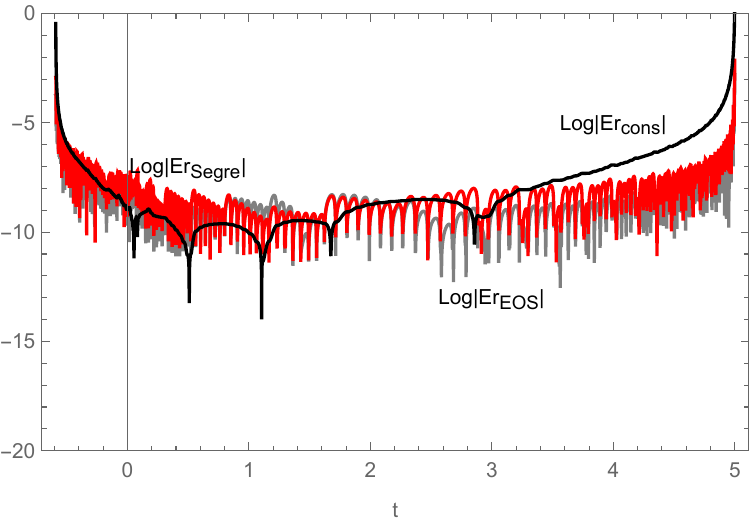}
    \includegraphics[width=7cm]{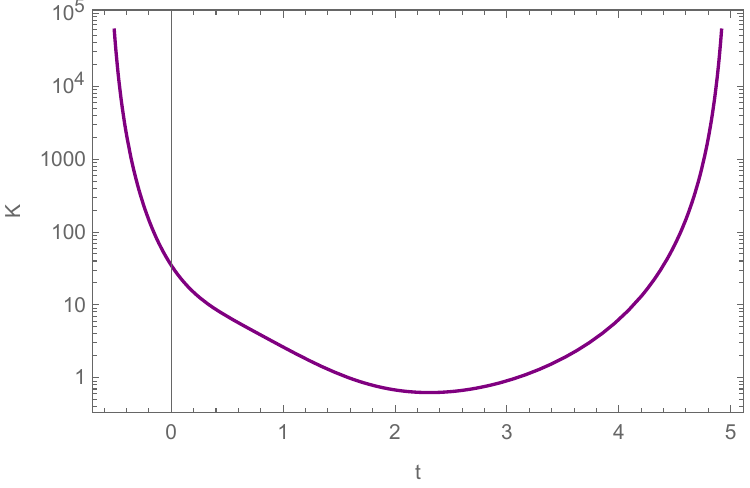}
    \caption{Plots of the scale functions $a,b$ (top left panel), $\rho=p_z=-p_\perp$ (top right panel), Error parameters \ref{errordefs} (bottom left panel), and Kretchman scalar (bottom right panel) for $k=1,\dot{b}=1,\ddot{b}=-1$. Unlike the corresponding $k=0$ case, the $b$ function collapses to $0$ and there is another singularity. However, the profile of the energy density is somewhat similar in that there is a peak shortly after $t=0$. }
    \label{fig:kp1}
\end{figure}

\begin{figure}[h]
    \centering
    \includegraphics[width=7cm]{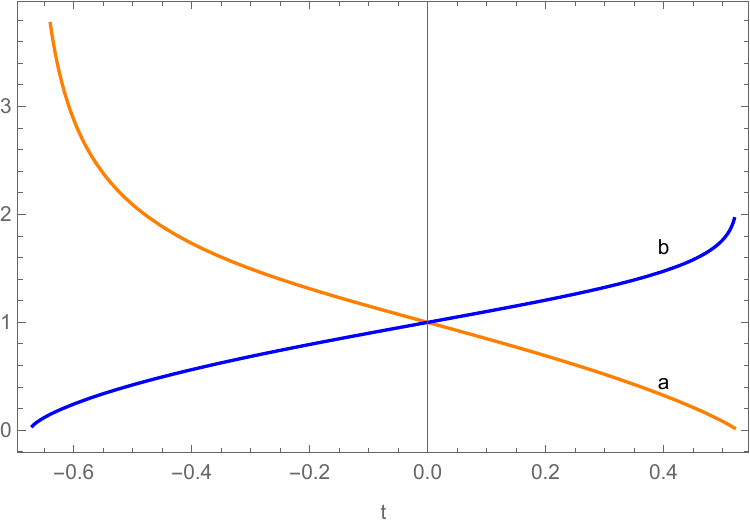}
    \includegraphics[width=7cm]{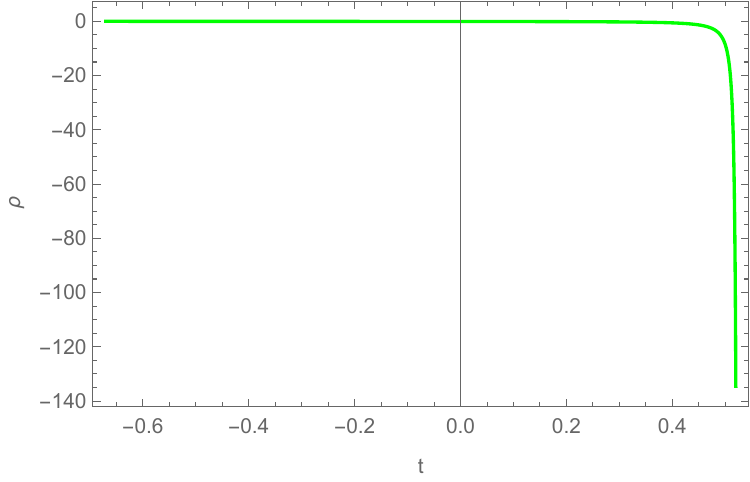}
    \includegraphics[width=7cm]{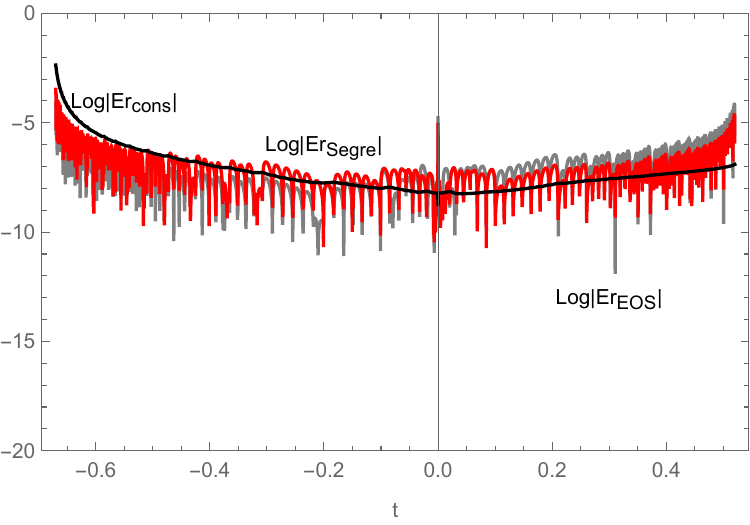}
    \includegraphics[width=7cm]{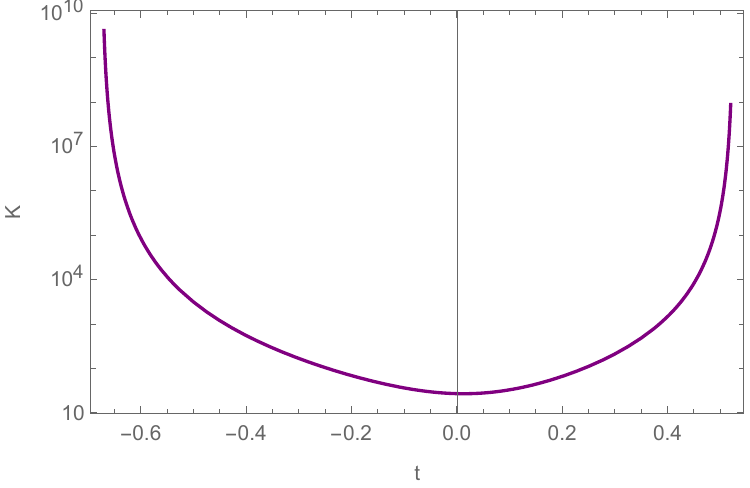}
    \caption{Plots of the scale functions $a,b$ (top left panel), $\rho=p_z=-p_\perp$ (top right panel), Error parameters \ref{errordefs} (bottom left panel), and Kretchman scalar (bottom right panel) for $k=1,\dot{b}=1,\ddot{b}=0$. This is rather similar to the $k=0$ case. The singularities are at $t\approx-0.67$ and $t\approx0.52$}
    \label{fig:kp2}
\end{figure}

\begin{figure}[h]
    \centering
    \includegraphics[width=7cm]{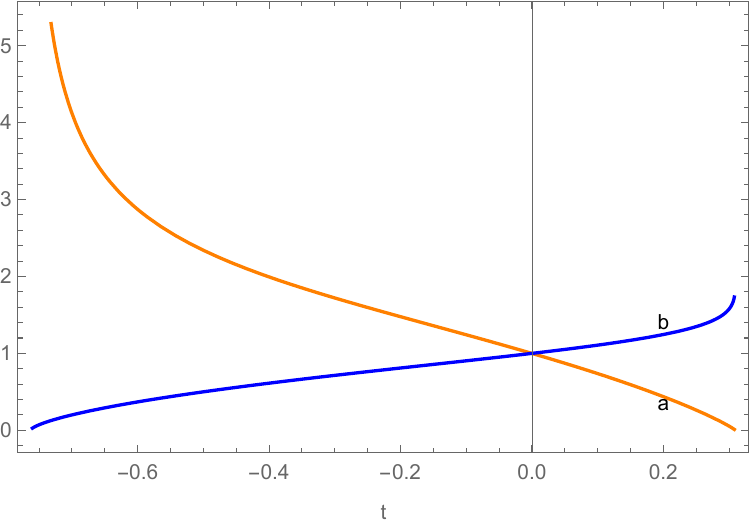}
    \includegraphics[width=7cm]{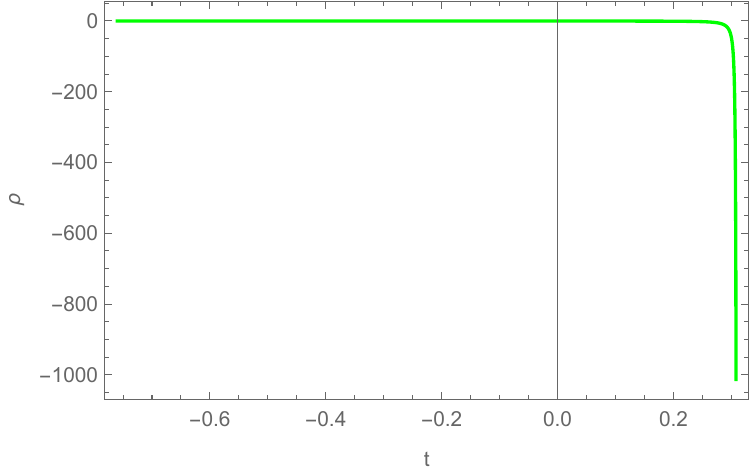}
    \includegraphics[width=7cm]{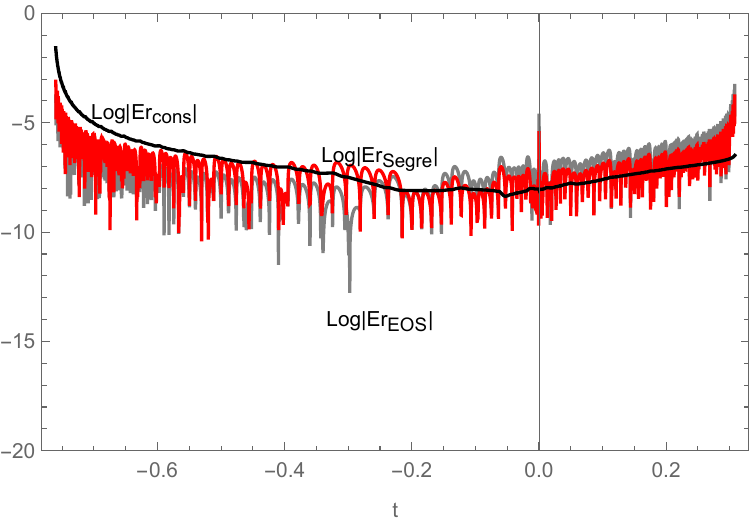}
    \includegraphics[width=7cm]{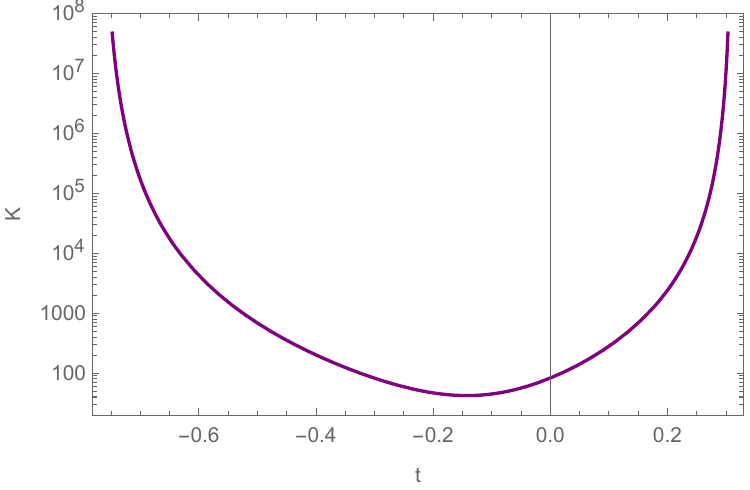}
    \caption{Plots of the scale functions $a,b$ (top left panel), $\rho=p_z=-p_\perp$ (top right panel), Error parameters \ref{errordefs} (bottom left panel), and Kretchman scalar (bottom right panel) for $k=1,\dot{b}=1,\ddot{b}=1$. This is rather similar to the $k=0$ case. The singularities are at $t\approx-0.76$ and $t\approx 0.31$.}
    \label{fig:kp3}
\end{figure}

We plot the $a,b$ scale functions, density, error parameters, and Kretchman scalar for the $\ddot{b}=-1$ in Figure \ref{fig:kp1}, $\ddot{b}=0$ in Figure \ref{fig:kp2}, and $\ddot{b}=1$ in Figure \ref{fig:kp3}. One additional noteworthy difference between the $k=0$ and $k=1$ calculations is that for $k=0$ the average magnitude of $Er_{EOS}$ was much lower than the other two while for $k=1$ they are more similar, this is likely due to the fact that we could use a simplified version of Eq. (\ref{anumB}) for $k=0$ but had to use Eq. (\ref{abcoupled}) here.
\subsubsection{k=-1}
\begin{figure}[h]
    \centering
    \includegraphics[width=7cm]{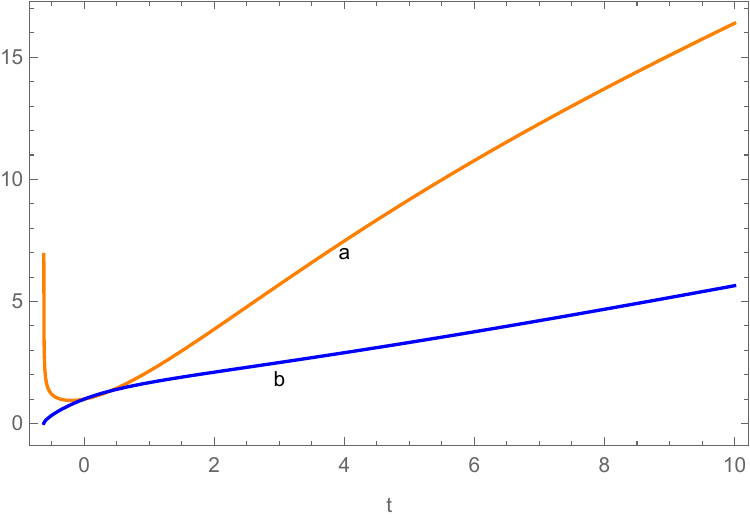}
    \includegraphics[width=7cm]{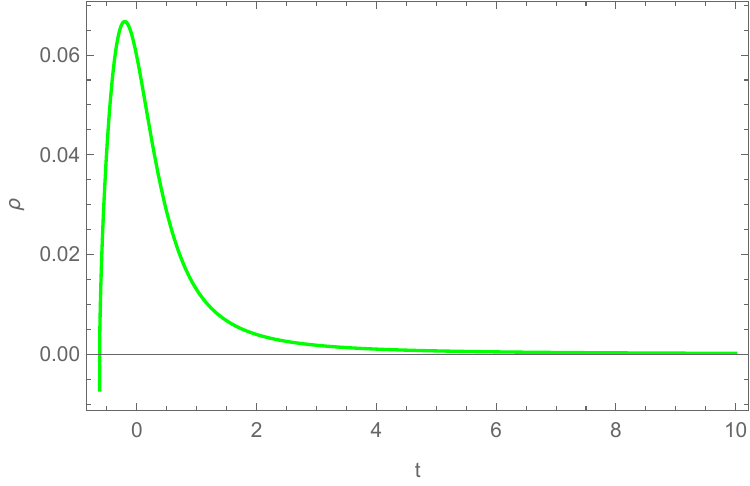}
    \includegraphics[width=7cm]{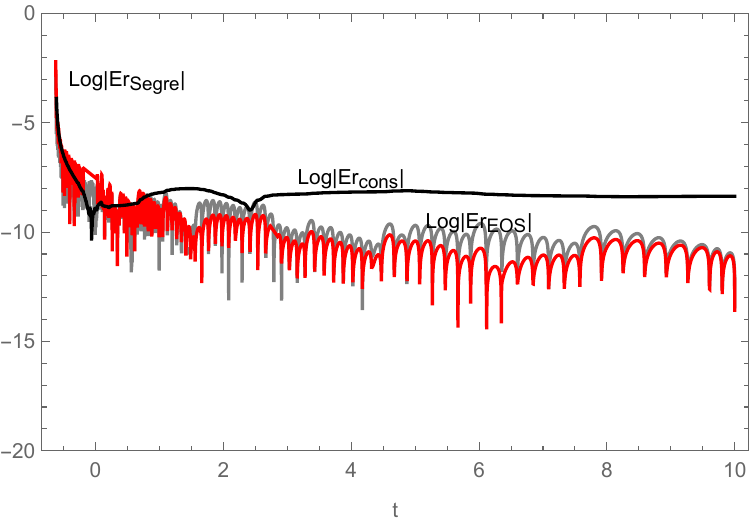}
    \includegraphics[width=7cm]{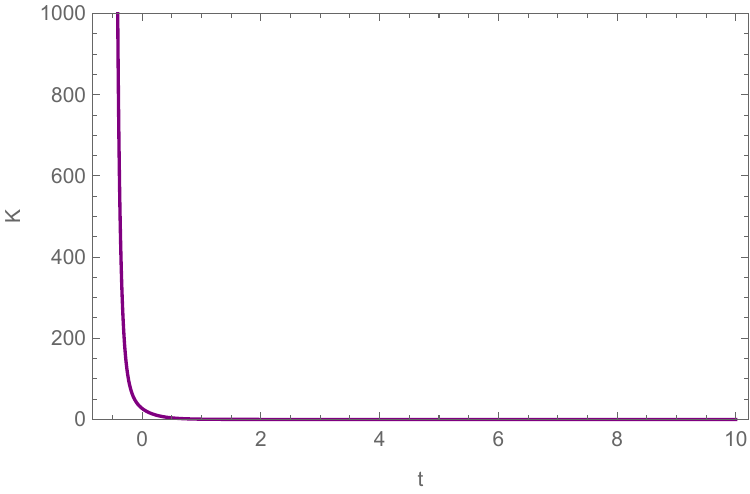}
    \caption{Plots of the scale functions $a,b$ (top left panel), $\rho=p_z=-p_\perp$ (top right panel), Error parameters \ref{errordefs} (bottom left panel), and Kretchman scalar (bottom right panel) for $k=-1,\dot{b}=1,\ddot{b}=-1$. We have qualitatively similar behavior to the $k=0,\ddot{b}=-1$ case Figure \ref{fig:kz1}. Our only bound is $t\approx-0.621.$ }
    \label{fig:km1}
\end{figure}

\begin{figure}[h]
    \centering
    \includegraphics[width=7cm]{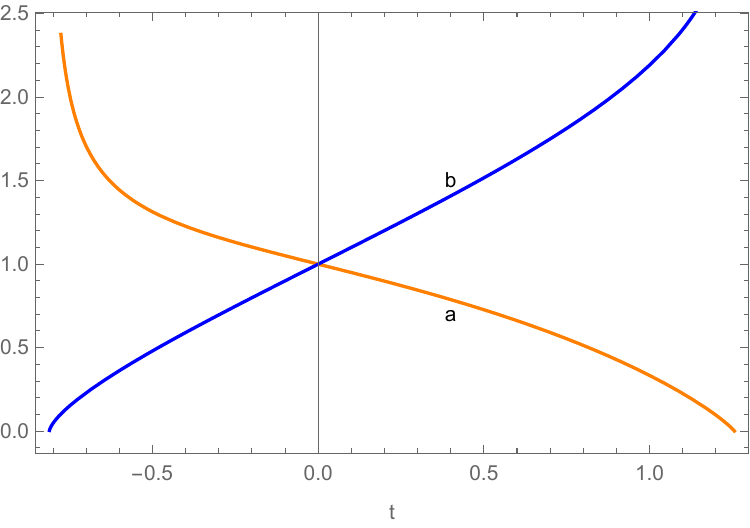}
    \includegraphics[width=7cm]{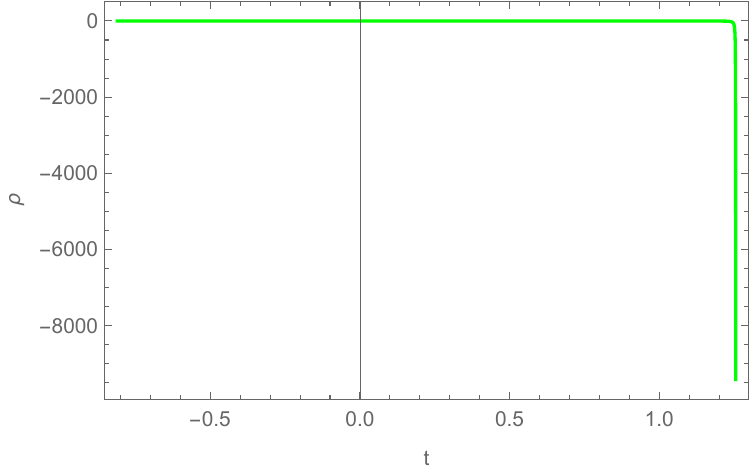}
    \includegraphics[width=7cm]{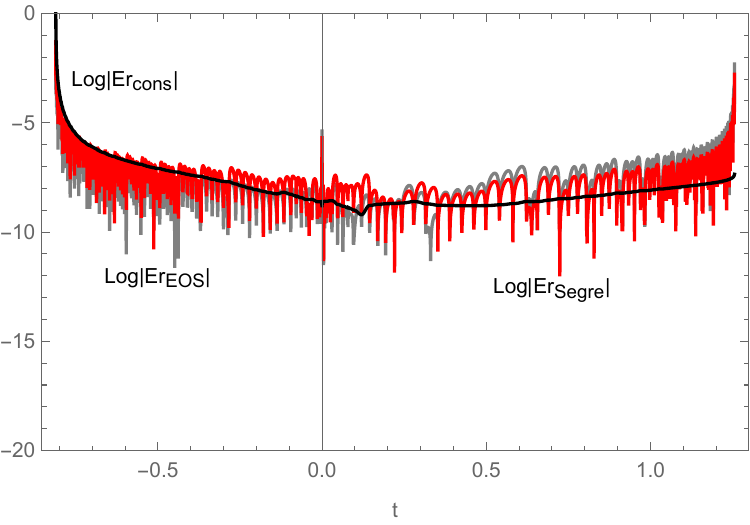}
    \includegraphics[width=7cm]{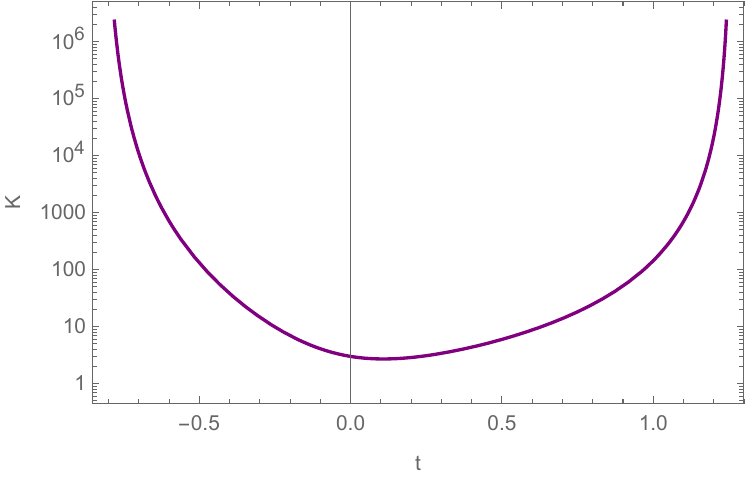}
    \caption{Plots of the scale functions $a,b$ (top left panel), $\rho=p_z=-p_\perp$ (top right panel), Error parameters \ref{errordefs} (bottom left panel), and Kretchman scalar (bottom right panel) for $k=-1,\dot{b}=1,\ddot{b}=0$. We are bounded on both sides by singularities at $t\approx-.811$, $t\approx 1.256$. This is qualitatively very similar to the other $\ddot{b}=0$ cases illustrated in Figures \ref{fig:kz2} and \ref{fig:kp2}. }
    \label{fig:km2}
\end{figure}

For the $k=-1$ cases, we also used a scheme based on Eqs. (\ref{abcoupled}) to compute the behavior for initial conditions of $b=1,\dot{b}=1,\ddot{b}=-1,0,1$, where once again switching $\dot{b}=-1$ gives the time reversed behavior.
From Figure \ref{fig:km1}, we see there are a lot of qualitative similarities between the $k=0$ and $k=-1$ behavior for $\ddot{b}=-1$. In both instances, we have $b\rightarrow0$ at a finite point in the past ($t\approx-0.621$ in this case), expansion of both $a$ and $b$ at large times, and a local minimum in $a$. Likewise, there is a local maximum in density and the density tapers off with expansion. The Kretchman scalar appears to diverge as we approach the $b=0$ point. There are however a few important differences. One difference is the large time expansion behavior of $b$ is much faster than in the $k=0$ case. Also, the error parameters have a rather different behavior in that the EOS error is more similar in magnitude to the others, but that is likely because the $k=-1$ needs to be implemented with Eq. (\ref{abcoupled}) rather than the simplified version of Eq. (\ref{anumB}).
The $\ddot{b}=0$ behavior is qualitatively similar to that for $k=0,1$, as we can see from Figure \ref{fig:km2}. Specifically, we start with a $b=0$ singularity with high Krechman scalar but low density, and end at with an $a=0$ singularity where both the Kretchman scalar and density are diverging. The density is again negative so the WEC is violated. The initial singularity is located at $t\approx-0.811$ and the final singularity is located at $t\approx 1.256$.
Finally, the $\ddot{b}=1,k=-1$ case has qualitatively new behavior. Like all the other cases except for $k=0,-1$ and $\ddot{b}=-1$, we are bounded by singularities on both sides, now at $t\approx-1.358$ and $t\approx 0.443$. However, in the initial singularity, both $a$ and $b$ go to 0, which is the only case we have examined where both scale functions have gone to 0. The second singularity has only $a$ going to 0. Because $a$ is going to 0 at both singularities, both singularities are associated with a divergence in density here. The density is still negative and the WEC is violated.
\begin{figure}[h]
    \centering
    \includegraphics[width=7cm]{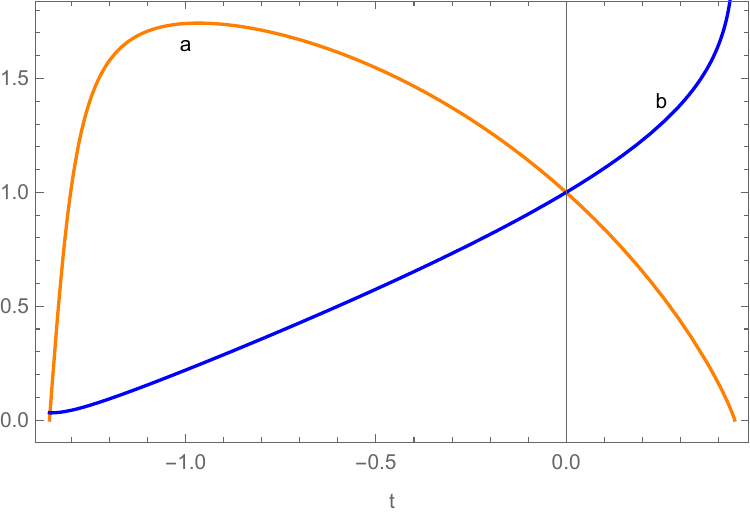}
    \includegraphics[width=7cm]{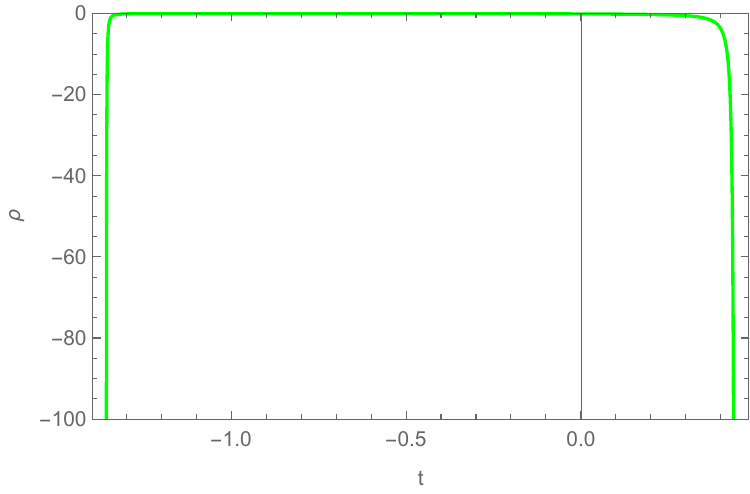}
    \includegraphics[width=7cm]{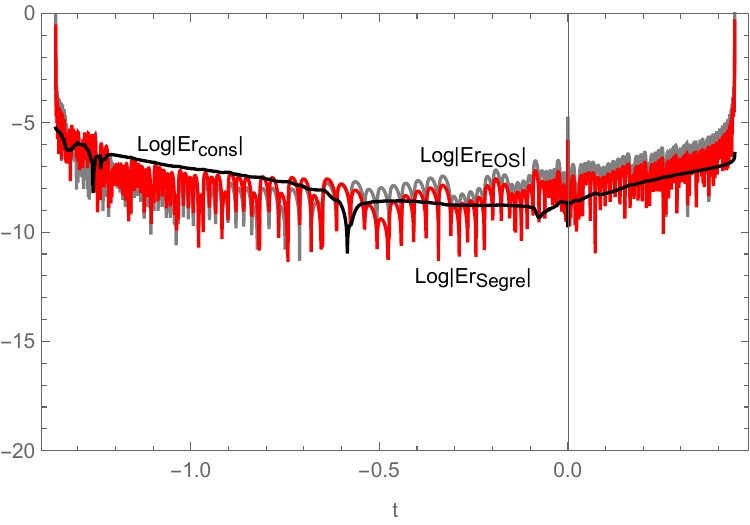}
    \includegraphics[width=7cm]{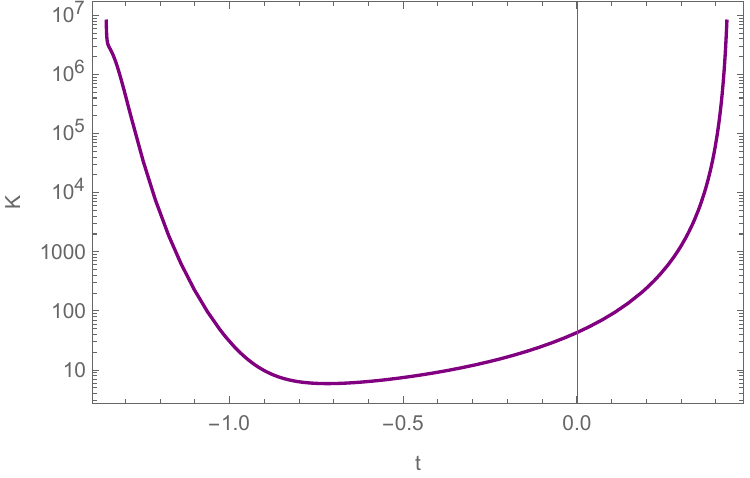}
    \caption{Plots of the scale functions $a,b$ (top left panel), $\rho=p_z=-p_\perp$ (top right panel), Error parameters \ref{errordefs} (bottom left panel), and Kretchman scalar (bottom right panel) for $k=-1,\dot{b}=1,\ddot{b}=1$. We have qualitatively new behavior compared to the previous cases in that $a\rightarrow 0$ at both singularities $t\approx-1.358$ and $t\approx 0.443$,}
    \label{fig:km3}
\end{figure}
\subsection{Summary of Qualitative Behaviors}
The most common behavior features a situation where we move between a singularity with $b=0$ and $a$ finite to a second singularity with $a=0$, $b$ finite, and negative infinite density. This occurs for all three cases where we chose the initial condition of $\ddot{b}=0$ ($k=0,1,-1$), as well as two of the three $\ddot{b}=1$ cases ($k=0,1$). 

The next most common behavior was an initial singularity in $b=0$, a local minimum in $a$ corresponding to a maximum in density, then continued expansion, occurring for the $\ddot{b}=-1$ initial condition for $k=0,-1$. The $\ddot{b}=-1,k=1$ case has similar behavior at early times, but collapses into a second $b=0$ singularity after finite time. Notice that these are the only situations where the density was positive (other than possibly very near the initial singularities, but that is likely numerical error).

Finally, the $\ddot{b}=1,k=-1$ case is unique in that it has $a=0$ at both the initial and final singularities, as well a a negative infinite density at both singularities. The initial singularity has $b=0$ but the final singularity seems to have finite $b$.

Notice that in all tested cases, switching the sign on the initial condition for $\dot{b}$ resulted in a time reversed situation, which agrees with the structure of Eqs. (\ref{abcoupled}) under time reversal ($\dot{a},\dot{b}$ themselves change signs but all terms in the equation take the same values).
\section{discussion and conclusion}
\label{conclusion}
In addition to vacuum energy and isotropic perfect fluids, there has been some interest in anisotropic models of dark energy. In this note we derive the equations for Segre [1(11,1)] spacetimes which obey the vacuum energy equation of state in a plane.

We use the metric \ref{met} because it is naturally adapted to describing spacetimes where isotropy is violated along a single axis. The metric is Bianchi type III when the spatial transverse curvature $k$ is nonzero and is Bianchi type I otherwise. 

We show the case when the distinct pressure follows the equation of state $p_\parallel=0$ there is a closed form solution parameterizing the time evolution of the spacetime. In the case that $p_\parallel=-\rho$ is applied, the symmetry degenerates into Segre type [(111,1)] and the solution to the Einstein equations becomes one of those previously presented in \cite{2023arXiv230109204B}, although in this paper we present additional cases in which Eq. (\ref{avack}) can be explicitly integrated. Finally, while we do not find any situations in which the equation of state $p_\parallel=\rho$ can be explicitly integrated, we perform numerical analysis to time evolve the state forwards and backwards for various initial conditions allowing for a qualitative picture of some of the behaviors such a system might manifest.

There are a few possible avenues for future work. One of which is to determine what matter fields result in Segre type [1(11,1)] energy-momentum tensors. While Segre type [(111,1)] can be interpreted as vacuum energy, and Segre type [(11)(1,1)] shows up in electrodynamics type theories, and [(111),1] is famously perfect fluid, we are not aware of systems with [1(11,1)], and an example system is not mentioned in the discussion of Segre type in \cite{Stephani:2003tm}.

\bibliography{monster.bib}

\end{document}